\documentclass{article}[12pt]
\usepackage{epsfig}

\newcommand{\beq}{\begin{equation}}
\newcommand{\eeq}{\end{equation}}
\newcommand{\beqn}{\begin{eqnarray}}
\newcommand{\eeqn}{\end{eqnarray}}
\newcommand{\bequ}{\begin{displaymath}}
\newcommand{\eequ}{\end{displaymath}}
\newcommand{\beqnu}{\begin{eqnarray*}}
\newcommand{\eeqnu}{\end{eqnarray*}}
\newcommand{\eqref}[1]{(\ref{#1})}

\newcommand{\muhat}{\hat\mu}
\newcommand{\nuhat}{\hat\nu}

\newcommand{\gbar}{\bar g} 
\newcommand{\csw}{c_{\rm sw}}
\newcommand{\cttilde}{\tilde{c}_t}

\newcommand{\rmO}{{\rm O}}
\newcommand{\Nf}{N_{\rm f}}

\newcommand{\Tr}{\mbox{Tr}}

\renewcommand{\Re}{\mbox{Re}}

\title{Schr\"odinger functional at negative flavour number}
\author{Bernd Gehrmann, Juri Rolf, Stefan Kurth, Ulli Wolff\\[5mm]
Institut f\"ur Physik, Humboldt-Universit\"at zu Berlin,\\
Invalidenstr. 110, 10115 Berlin, Germany}

\begin{document}

\addtolength{\topmargin}{-2\baselineskip}
\addtolength{\textheight}{4\baselineskip}

\maketitle
\centerline{
\includegraphics[width=2.5cm]{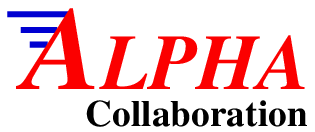}
}
\vskip 0.5 cm

\begin{abstract}

The scaling of the Schr\"odinger functional coupling is studied numerically and
perturbatively for an SU(3) lattice gauge field coupled to an $O(a)$
improved bosonic spinor field.  This corresponds to QCD with minus two
light flavours and is used as a numerically less costly test case for
real QCD.  A suitable algorithm is developed, and the influence of the
matter fields on the continuum limit and the lattice artefacts are
studied in detail.
\end{abstract}

\section{Introduction}

The strong coupling constant $\alpha_s$ of QCD is of particular theoretical
interest. On the one hand, it can be extracted from jet events which
are a property of the strong interaction at large energies. On the
other hand, as has been discussed in detail in~\cite{Weisz:1995yz}, the running
coupling can be computed in lattice gauge theory. There, the
parameters may be fixed in the non perturbative hadronic regime,
taking as experimental input for example the pion decay constant and
the masses of the $\pi,\ K,\ D$ and $B$. The computation of
the running of the coupling up to large energies thus provides a
quantitative test of the theory, which is believed to be fundamental 
in the hadronic as well as in the high energy
regime. Furthermore, it is interesting to find out at which energies
the perturbative behaviour of a given coupling sets in.

The basic strategy for such a computation has been proposed by
L\"uscher, Weisz and Wolff~\cite{Luscher:1991wu}. They use a non perturbative
definition of the coupling, which runs with the spacetime volume. Its
evolution is mapped out
by a recursive finite size scaling technique up to large
energies where contact with the
minimal subtraction scheme is made by perturbation theory.
The central object in this computation is
the step scaling function, which can be understood as a
beta function for finite scale transformations.
At each step in the recursive evolution of the coupling,
the continuum limit is taken. Other key ingredients include
$\rmO(a)$ improvement and Schr\"odinger functional boundary
conditions~\cite{heplat9207009,Sint:1994un,heplat9609035}.

This strategy applies to any asymptotically free theory,
and it has first been tested for the nonlinear $O(3)$ model in two
dimensions~\cite{Luscher:1991wu}, pure $SU(2)$ gauge theory~\cite{deDivitiis:1995yz}, 
and pure $SU(3)$
gauge theory~\cite{heplat9309005,Garden:1999fg}, 
which can also be interpreted as the quenched
approximation of QCD. In~\cite{Bode:2001jv}, the ALPHA collaboration has just
published their first quantitative results for the evolution of the
coupling in QCD with two flavours. 

However, since simulations in full QCD are notorious for high
computational cost, the data at $\Nf=2$ do not yet reach very close to
the continuum limit in the individual steps of the
computation. Therefore, we have decided to also study the approach to
the continuum in a simpler model that is more easily accessible to
simulation.  It differs from the quenched approximation by depending
on the fermionic determinant. In this computation, the focus is not so
much on the running coupling as a function of the energy scale, but
rather on the details of the approach of the continuum limit in the
perturbative regime and at slightly lower energy.

The model we investigate in this paper may be viewed as arising from
an analytic continuation of the flavour number to negative values and
in particular to $\Nf=-2$.  Since a negative power of the fermionic
determinant may be represented by bosonic spinor fields with the same
indices as fermionic fields, the name \emph{bermions} was coined for
such theories~\cite{heplat9507020}.  The main virtue of these models
is that the interaction term becomes local and thus numerical
simulations are considerably cheaper than in full QCD.

In the literature (e.g.~\cite{Anthony:1982fe,heplat9507020}), they were
mainly considered from an algorithmic point of view with the idea in
mind to extrapolate in $\Nf$ from negative values to $\Nf=2$.  
However, this is problematical, since fermionic zero modes may be
encountered (for example at small quark masses or in large physical
volume), which dominate the dynamics in the theories at negative
flavour numbers. 
Thus,  in our work, no extrapolation of
results from negative to positive values of $\Nf$ is attempted or
aimed at.

In~\cite{heplat9907007}, two of the authors have published results for
the step scaling function for unimproved Wilson bermions ($\Nf=-2$) in
the perturbative regime. Lattice artefacts turned out to be very
large.  As for the quenched approximation and for full QCD with two
flavours, we now study the $\rmO(a)$ improved $\Nf=-2$ theory. The
inclusion of the clover term into the bermionic action poses certain
algorithmic problems that are dealt with in this article. We present a
detailed study of the performance of the algorithm used for our Monte
Carlo simulations.

An important input for the understanding of our Monte Carlo data comes
from lattice perturbation theory. The cutoff effects of the step
scaling function can be computed perturbatively. They are used in the
data analysis. The cutoff effects have already been estimated
in~\cite{heplat9911018} to 2-loop order.  However, in that calculation,
the continuum value of the critical mass has been used as a first
estimate instead of the finite lattice value, which was not yet
available (see discussion in ~\cite{heplat9911018}). The computation
of this critical mass is technically more involved due to extra
tadpole diagrams that emerge from the non vanishing background
field. Here we present a computation that includes all diagrams and
completes the study of~\cite{heplat9911018}.

This article is organized in the following way. In the next section, we
reflect the most important definitions that occurred in our previous
articles. In section~3, we discuss the perturbative expansion of the
lattice artefacts of the step scaling function. After that, the
bermion model and the algorithm used in our non-perturbative
calculations is discussed. In the last section, our numerical results
are summarized.

\section{Lattice theory}
\label{sec:latticetheory}

Since this work extends our earlier work reported
in~\cite{heplat9207009,heplat9309005,heplat9907007}, we only briefly
summarize the necessary notations. For unexplained conventions, we
refer in particular to~\cite{heplat9605038,heplat9606016}.

The theory is set up on a four dimensional hyper-cubic Euclidean
lattice with lattice spacing $a$ and size $T\times L^3, T=L$, $L$
being an integer multiple of $a$. The gauge field on this lattice is
represented by an $SU(3)$ matrix $U(x,\mu)$ that is defined on every
link between nearest neighbour sites $x$ and $x+a\hat{\mu}$ of the
lattice ($\hat{\mu}$ denotes the unit vector in the direction
$\mu=0,1,2,3$). Furthermore, on the lattice sites reside $\Nf$
flavours of mass degenerate fermionic quark fields $\psi(x)$, which
also carry Dirac and colour indices. We do not specify $\Nf$ at the
moment. Later, we want to consider the theory in which $\Nf$ is
continued to negative numbers. This has to be done after
the integration over the quark fields has been performed.

The spatial sub-lattices at fixed times $x_0$ are thought to be
wrapped on a torus. The gauge field thus fulfils periodic boundary
conditions in the space directions while the quark fields obey
periodic boundary conditions in these directions up to a phase
$e^{i\theta}$~\cite{Sint:1996ch}. In the time direction, we impose
Dirichlet boundary conditions. The gauge field at the boundary takes
the form
\beqn 
  U(x,k)|_{x_0=0} &=& \exp(aC), \nonumber\\
  U(x,k)|_{x_0=T} &=& \exp(aC').  
\eeqn
The constant diagonal fields $C$ and $C'$ can be chosen such that a
constant colour electric background field is enforced on the
system~\cite{heplat9207009}.  The boundary conditions for the quark
fields are discussed in detail in~\cite{heplat9606016}. The boundary
quark fields serve as sources for fermionic correlation
functions. They are set to zero after differentiation.

The Schr\"odinger functional is the partition function of the system,
\beq
  Z = e^{-\Gamma} = \int\! D[U]D[\bar\psi]D[\psi] \, e^{-S[U,\bar\psi,\psi]}.
\eeq
It involves an integration over the fields with fixed boundary values
at $x_0=0$ and $x_0=T$. For the action, we take the sum
\beq
  S[U,\bar\psi,\psi] = S_g[U] + S_f[U,\bar\psi,\psi]
\eeq
of the $\rmO(a)$ improved plaquette action
\beqn
  S_g[U] = \frac{1}{g_0^2} \sum_p w(p) {\rm tr}(1-U(p))
\eeqn
and the fermionic action
\beqn
  S_f[U,\bar\psi,\psi] = \sum_x \bar\psi(x) (D+m_0) \psi(x).
\eeqn
Here, $D$ is the $\rmO(a)$ improved Wilson Dirac operator including the
Sheikholeslami-Wohlert term~\cite{Sheikholeslami:1985ij} multiplied
with the improvement coefficient $\csw$ and a boundary improvement
term that goes with $\cttilde$.  Details can be found
in~\cite{heplat9605038,heplat9606016}.  As discussed for example
in~\cite{heplat9207009}, the leading cutoff effects from the gauge
action can be cancelled by adjusting the weights $w(p)$ of the
plaquettes at the boundary, i.e. one sets \beqn w(p) = c_t(g_0) \eeqn
if $p$ is a time-like plaquette attached to a boundary plane. In all
other cases, $w(p)=1$.

The improvement coefficient $\csw$ has been computed to
1-loop order of perturbation theory with the 
result~\cite{Wohlert:1987rf,heplat9606016}
\beqn
  \csw(g_0) = 1 + 0.26590(7) g_0^2,
\eeqn
independent of $\Nf$ to this order.
For $\Nf=0$ and $\Nf=2,\ $ $\csw$ has also been computed
non-perturbatively~\cite{heplat9609035,heplat9709022}. The results of
these simulations can be represented in the region $0 \le g_0 \le 1$ 
in good approximation by the rational functions 
\beqn
  \csw(g_0)\big|_{\Nf=0} &=& 
    \frac{1-0.656 g_0^2-0.152 g_0^4-0.054 g_0^6}
         {1-0.922 g_0^2},  \nonumber\\
  \csw(g_0)\big|_{\Nf=2} &=&
    \frac{1-0.454 g_0^2-0.175 g_0^4+0.012 g_0^6+0.045 g_0^8}
         {1-0.720 g_0^2}. 
\eeqn
The boundary improvement coefficients are only known
perturbatively. The 2-loop value for $c_t$ depends 
(in principle) quadratically on
$\Nf$ and has the form
\beqn
  c_t(g_0) &=& 1+\big( -0.08900(5) + 0.0191410(1) \Nf\big) g_0^2 \nonumber\\
           &&\quad + \big( -0.0294(3) + 0.002(1) \Nf + 0.0000(1) \Nf^2\big) g_0^4.
\eeqn
$\cttilde$ is known to 1-loop order,
\beqn
  \tilde c_t(g_0) = 1 -0.0180(1) g_0^2.
\eeqn

From the Schr\"odinger functional, a running coupling may be defined by
differentiating with respect to the boundary fields. To obtain a
complete definition, the diagonal matrices $C$ and $C'$ and the
direction of the differentiation must be specified. Here we
differentiate along a curve parametrized by the dimensionless
parameter $\eta$ at the boundary field "A" of
reference~\cite{heplat9309005}, which is favoured by practical
considerations such as mild cutoff effects.  With this choice, the
induced constant colour electric background field can be represented
by
\beqn
  V(x,\mu) = e^{aB_{\mu}(x)},
\eeqn
with
\beqn
  B_0 = 0, \qquad B_k = \big(x_0 C' + (T-x_0)C\big)/T.
\eeqn
Now, since $\Gamma'=-\frac{\partial\log Z}{\partial\eta}$ is a
renormalized quantity~\cite{heplat9309005} with the perturbative
expansion $\Gamma'=g_0^{-2}\Gamma_0'+\Gamma_1'+\ldots$, a renormalized
coupling with the correct normalization is defined as
\beqn
  \bar g^2(L) = \left. \frac{\Gamma_0'}{\Gamma'} \right|_{\eta=0}.
\eeqn
This coupling can be computed efficiently in numerical
simulations as the expectation value
$\frac{\partial\Gamma}{\partial\eta} = \left\langle 
  \frac{\partial S}{\partial\eta} \right\rangle$. 

For $\Nf\neq 0$, the coupling depends not only on the scale $L$ but
also on the mass $m_1$ of the quarks, which we define via the PCAC
relation~\cite{Jansen:1996ck}. To this end, the fermionic boundary
fields of the Schr\"odinger functional are used to transform this
operator relation to an identity that holds up to $\rmO(a^2)$ between
improved fermionic correlation functions on the lattice.  In
section~\ref{sec:pert}, this will be explained in more detail.

To define the step scaling function $\sigma(u)$, we set $u=\gbar^2(L)$
and tune $m_1(L/a)=0$. Then we change the length scale by a factor $2$ and
compute the new coupling $u'=\gbar^2(2L)$. The lattice step scaling
function $\Sigma$ at resolution $L/a$ is defined as
\beqn
  \Sigma(u,a/L) = \left. \gbar(2L)\right\vert_{u=\gbar^2(L), m_1(L/a)=0}.
\eeqn
These conditions on $\gbar^2$ and $m_1$ fix the bare
parameters of the theory. The continuum limit $\sigma(u)$ can be
found by an extrapolation in $a/L$. We expect that in the $\rmO(a)$
improved theory $\Sigma(u,a/L)$ converges to $\sigma(u)$ with a
rate roughly proportional (i.e. up to logarithms and higher orders) 
to $(a/L)^2$.

\section{Perturbative computation of the cutoff effects}
\label{sec:pert}

The size of the cutoff effects in the step scaling function can be
estimated in perturbation theory. To this end, the relative deviation
of the step scaling function from its continuum limit is expanded in
powers of $u$,
\beqn
  \delta(u,a/L) &=& \frac{\Sigma(u,a/L)-\sigma(u)}{\sigma(u)}\nonumber \\
  &=&  
   \left[\delta_{10} + \delta_{11}\Nf\right]u
 + \left[\delta_{20}+\delta_{21}\Nf+\delta_{22}\Nf^2\right]u^2 
 + \rmO(u^3).
\label{eq:delta}
\eeqn
It turns out to be quite small at 2-loop level.  However, it is still
necessary to extrapolate the Monte Carlo data to the continuum limit
by simulating a sequence of lattice pairs with decreasing lattice
spacing and fixed coupling $u$.  One may use the perturbative
expansion of $\delta(u,a/L)$ to remove the $\rmO(a)$ cutoff effects up to
2-loop order from the non-perturbative values of the step scaling
function $\Sigma(u,a/L)$.

The 1-loop coefficients $\delta_{1j}(a/L)$,
first listed in \cite{sommerunpublished},  are shown in
table~\ref{tab:pertresults}.  The 2-loop coefficients $\delta_{2j}(a/L)$
have been estimated in~\cite{heplat9911018}.  However, the parts of
$\delta_{2j}$ involving contributions from the quarks contain the 1-loop
coefficient of the critical bare quark mass $m_c$ at which the renormalized
quark mass vanishes. This zero mass condition has to be specified with
the cutoff in place. Thus, in the expansion
\beqn 
m_c = m_c^{(0)} + m_c^{(1)}g_0^2 + \rmO(g_0^4),
\eeqn
we get $a/L$ dependent expansion coefficients
$m_c^{(i)}$, which only in the limit \mbox{$a/L\rightarrow 0$} go over to
their continuum values, which is $m_c^{(0)}=0$ at tree level, while
the 1-loop value $m_c^{(1)}$ can be found in table~1
of~\cite{heplat9911018}.  In~\cite{heplat9911018}, these continuum
values were used to compute the 2-loop coefficients $\delta_{2j}$. So, as
the authors state, the results presented there can only give a first
idea of the size of the cutoff effects. To obtain the correct values,
we need to compute $m_c^{(1)}$ at finite $a/L$.

\begin{table}[htbp]
  \begin{center}
    \begin{tabular}{rlllll}
      \hline\hline
      $L/a$ & $\delta_{10}$ & $\delta_{11}$ & 
     $\delta_{20}$ & $\delta_{21}$ & $\delta_{22}$ \\
      \hline
  4 & -0.01033 & $\,\!$ 0.00002 & -0.00159 & -0.00069 & 0.000724 \\
  5 & -0.00625 & -0.00013 & -0.00087 & -0.00048 & 0.000411 \\
  6 & -0.00394 & -0.00014 & -0.00055 & -0.00033 & 0.000199 \\
  7 & -0.00268 & -0.00014 & -0.00038 & -0.00021 & 0.000102 \\
  8 & -0.00194 & -0.00011 & -0.00027 & -0.00013 & 0.000058 \\
  9 & -0.00148 & -0.00009 & -0.00020 & -0.00010 & 0.000038 \\
 10 & -0.00117 & -0.00007 & -0.00015 & -0.00007 & 0.000026 \\
 11 & -0.00095 & -0.00006 & -0.00011 & -0.00006 & 0.000020 \\
 12 & -0.00079 & -0.00005 & -0.00009 & -0.00005 & 0.000016 \\
      \hline\hline
    \end{tabular}
    \caption{Perturbative results for $\delta(u,a/L)$ up to 2--loop order.}
    \label{tab:pertresults}
  \end{center}
\end{table}

For the quark mass, we adopt the definition of~\cite{heplat9605038}
based on the PCAC relation.  We first introduce the bare correlation
functions
\beqn
f_A(x_0) &=& -\frac{a^9}{L^3}\sum_{\mathbf{x},\mathbf{y},\mathbf{z}}\frac{1}{3}
\left\langle A_0^a(x)\bar{\zeta}(\mathbf{y})\gamma_5\frac{1}{2}
\tau^a\zeta(\mathbf{z})\right\rangle,\\
f_P(x_0) &=& -\frac{a^9}{L^3}\sum_{\mathbf{x},\mathbf{y},\mathbf{z}}\frac{1}{3}
\left\langle P^a(x)\bar{\zeta}(\mathbf{y})\gamma_5\frac{1}{2}
\tau^a\zeta(\mathbf{z})\right\rangle,
\eeqn
where $A^a$ and $P^a$ denote the axial current and density and $\zeta(\mathbf{x})$
is the functional derivative with respect to the boundary quark fields at
$x_0=0$. Now we can define the $x_0$ dependent current quark mass
\beq
m(x_0) = \frac{\frac{1}{2}(\partial_0^* + \partial_0)f_A(x_0)
 + c_Aa\partial_0^*\partial_0f_P(x_0)}{2f_P(x_0)},
\eeq
where $\partial_0$ and $\partial_0^*$ are the naive forward and
backward derivatives on the lattice.  As an unrenormalized quark mass,
we will use $m$ in the middle of the lattice, i.e.
\beq
  m_1 = \left\{ \begin{array}{ll}
      m\left(\frac T 2\right) 
      & \mbox{for even $T/a$,} \\
      \frac 1 2 \left( m\left(\frac{T-a}{2}\right)
                      +m\left(\frac{T+a}{2}\right)\right)
      & \mbox{for odd $T/a$.}
      \end{array} \right.
\eeq
The $\rmO(a)$ correction of the axial current is proportional to the
improvement coefficient $c_A$ which is
\beq
  c_A(g_0) = -0.00756(1) g_0^2
\eeq
to 1-loop order in perturbation theory~\cite{heplat9606016}.
Here, we are interested in the critical bare quark mass $m_c$ at
which the renormalized quark mass is zero.  Since $m_1$ is only
renormalized multiplicatively, it is sufficient to require $m_1=0$ in
order to make the renormalized quark mass vanish.

The current quark mass may be expanded in powers of $g_0^2$,
\beq
m_1 = m_1^{(0)}(m_0) + m_1^{(1)}(m_0)g_0^2 + \rmO(g_0^4),
\eeq
where the expansion coefficients 
depend on the bare quark mass $m_0$.
To set up perturbation theory, we consider $m_0$ also as a series,
\beq
m_0 = m_0^{(0)} + m_0^{(1)}g_0^2 + \rmO(g_0^4),
\eeq
and expand $m_1$ further as
\beq
m_1 = m_1^{(0)}\left(m_0^{(0)}\right) + \Biggl[m_1^{(1)}\left(m_0^{(0)}\right)
 + m_0^{(1)}\frac{\partial}{\partial m_0}m_1^{(0)}\left(m_0^{(0)}\right)
\Biggr]g_0^2 + \rmO(g_0^4).
\eeq
Therefore, the computation of $m_c^{(1)}$ has to be done in two steps. First we
compute $m_c^{(0)}$ by requiring
\beq
m_1^{(0)}\left(m_c^{(0)}\right) = 0,
\eeq
then we can determine $m_c^{(1)}$ from
\beq
m_1^{(1)}\left(m_c^{(0)}\right) 
+ m_c^{(1)}\frac{\partial}{\partial m_0}m_1^{(0)}\left(m_c^{(0)}\right) = 0.
\eeq
The first step is easily done numerically, the results are shown in
table~\ref{tab:pertmass}. The second step mainly amounts to expanding $f_A$ and
$f_P$ up to order $g_0^2$ and requires a slightly bigger effort.

\begin{table}[htbp]
  \begin{center}
    \begin{tabular}{rlll}
      \hline\hline
      $L/a$ & $m_c^{(0)}$ & $ m_{c0}^{(1)}$ & $m_{c1}^{(1)}$ \\
      \hline
  4 & -0.0015131 & -0.26667 & 0.0027136 \\
  5 & -0.0016969 & -0.26782 & 0.0006933 \\
  6 & -0.0006384 & -0.26984 & 0.0001990 \\
  7 & -0.0005761 & -0.26995 & 0.0000852 \\
  8 & -0.0003209 & -0.27004 & 0.0000451 \\
  9 & -0.0002753 & -0.27005 & 0.0000026 \\
 10 & -0.0001835 & -0.27006 & 0.0000017 \\
 11 & -0.0001561 & -0.27006 & 0.0000011 \\
 12 & -0.0001145 & -0.27007 & 0.0000008 \\
      \hline\hline
    \end{tabular}
    \caption{Perturbative results for the critical quark mass $m_c$ 
    up to 1--loop order.}
    \label{tab:pertmass}
  \end{center}
\end{table}

The expansion of $f_A$ and $f_P$ is outlined in~\cite{heplat9606016}.
After integrating over the quark fields and inserting
the contractions of the quark and boundary fields, one obtains
\beqn
f_{A,P}(x_0) &=& \cttilde^2\frac{a^9}{L^3}
\sum_{\mathbf{x},\mathbf{y},\mathbf{z}}\frac{1}{2}
\Bigl\langle \mbox{tr}\{P_+\Gamma P_- U(z-a\hat{0},0)S(z,x) \nonumber\\
& & \;\;\;\;\;\left. \Gamma S(x,y)U(y-a\hat{0},0)^{-1}
\}\Bigr\rangle_G\right|_{y_0=z_0=a},
\label{eq:fx}
\eeqn
where $\Gamma=\gamma_0\gamma_5$ for $f_A$ and $\Gamma=\gamma_5$
for $f_P$, and the trace is to be taken over the Dirac and colour indices only.
$P_+$ and $P_-$ are the projectors
\beq
P_\pm =\frac{1}{2}(1 \pm \gamma_0),
\eeq
and $\langle\ldots\rangle_G$ denotes the gauge field average with a
probability density proportional to
\beq
\det(D+m_0)\exp\{S_G[U]\}.
\eeq
The correlation functions $f_A$ and $f_P$ in (\ref{eq:fx})
contain two quantities that have to be expanded in powers of $g_0$.
One is the quark propagator
\beq
S(x,y) = S^{(0)}(x,y) + S^{(1)}(x,y)g_0 + S^{(2)}(x,y)g_0^2 + \rmO(g_0^3),
\eeq
the other one is the gauge field $U$, which is expanded around the
background field $V$
\beqn
U(x,\mu) &=& V(x,\mu)\exp\{g_0aq_{\mu}(x)\}\nonumber\\ 
&=&
V(x,\mu)\left(1 + g_0aq_{\mu}(x) + g_0^2a^2q_{\mu}(x)^2 + \rmO(g_0^3)\right).
\eeqn

\begin{figure}
  \noindent
  \begin{center}
  \begin{minipage}[b]{.3\linewidth}
     \centering\epsfig{figure=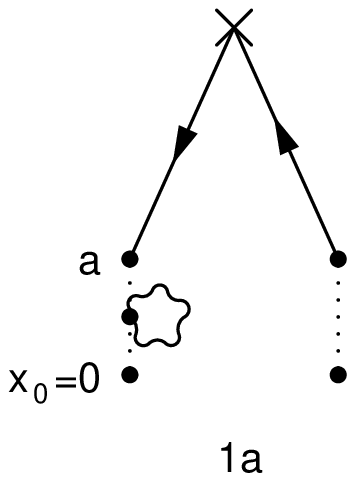,width=.8\linewidth}
  \end{minipage}
  \begin{minipage}[b]{.3\linewidth}
     \centering\epsfig{figure=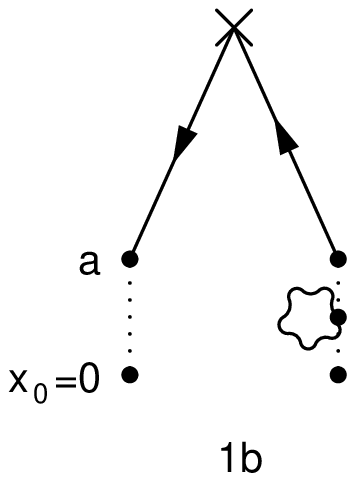,width=.8\linewidth}
  \end{minipage}
  \begin{minipage}[b]{.3\linewidth}
     \centering\epsfig{figure=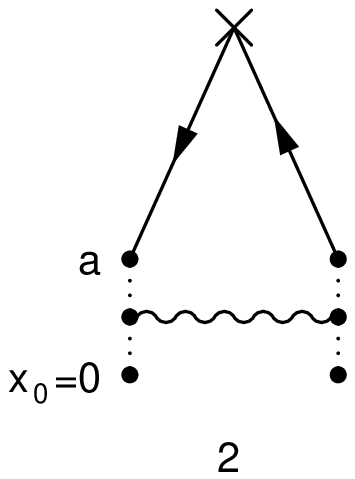,width=.8\linewidth}
  \end{minipage}\\
  \begin{minipage}[b]{.3\linewidth}
     \centering\epsfig{figure=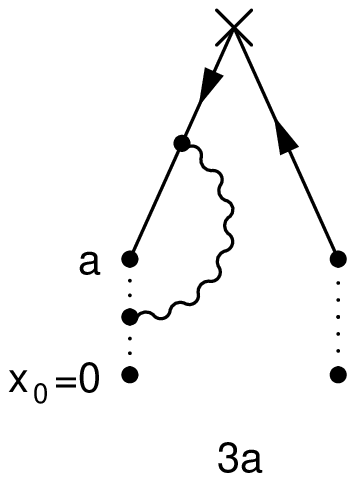,width=.8\linewidth}
  \end{minipage}
  \begin{minipage}[b]{.3\linewidth}
     \centering\epsfig{figure=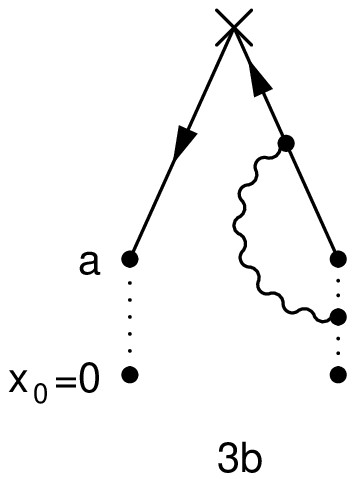,width=.8\linewidth}
  \end{minipage}
  \begin{minipage}[b]{.3\linewidth}
     \centering\epsfig{figure=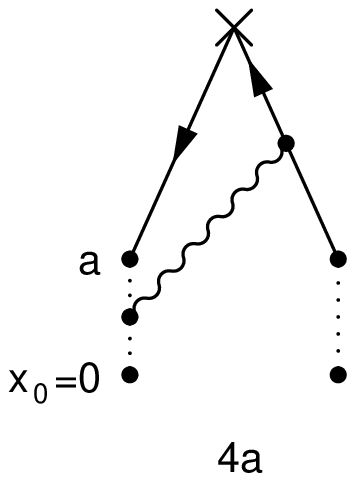,width=.8\linewidth}
  \end{minipage}\\
  \begin{minipage}[b]{.3\linewidth}
     \centering\epsfig{figure=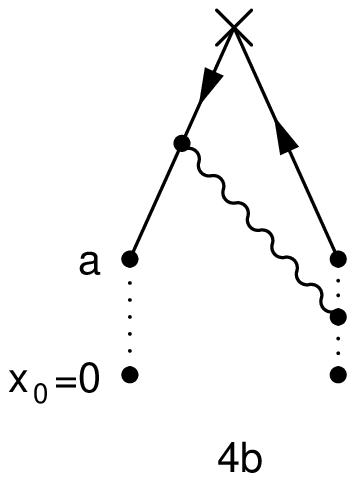,width=.8\linewidth}
  \end{minipage}
  \begin{minipage}[b]{.3\linewidth}
     \centering\epsfig{figure=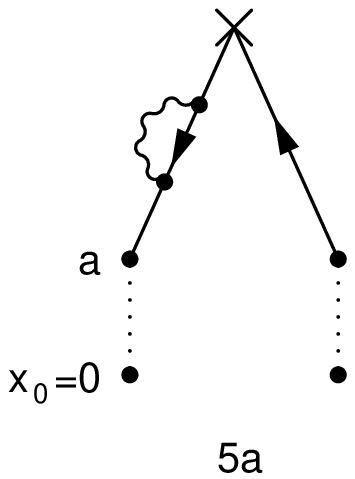,width=.8\linewidth}
  \end{minipage}
  \begin{minipage}[b]{.3\linewidth}
     \centering\epsfig{figure=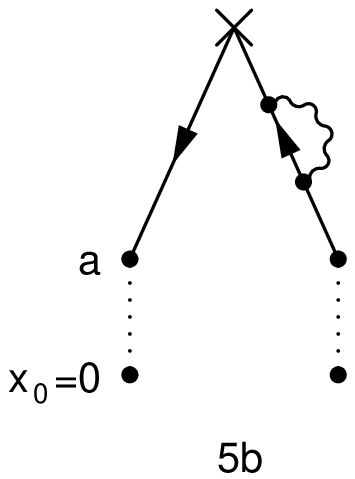,width=.8\linewidth}
  \end{minipage}\\
  \begin{minipage}[b]{.3\linewidth}
     \centering\epsfig{figure=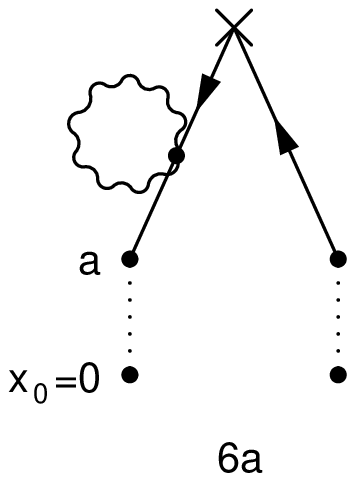,width=.8\linewidth}
  \end{minipage}
  \begin{minipage}[b]{.3\linewidth}
     \centering\epsfig{figure=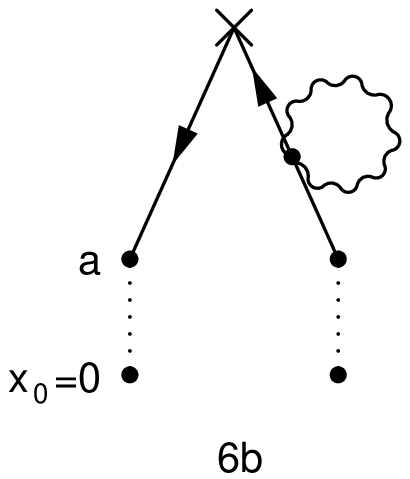,width=.8\linewidth}
  \end{minipage}
  \begin{minipage}[b]{.3\linewidth}
     \centering\epsfig{figure=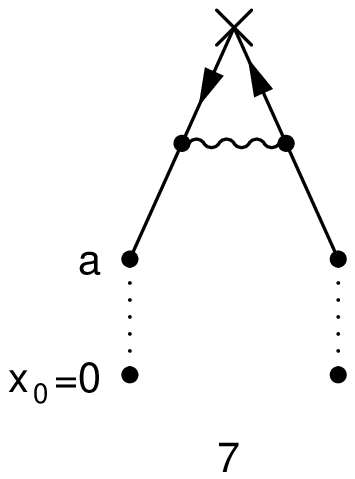,width=.8\linewidth}
  \end{minipage}
  \end{center}
  \caption{\label{fig:diagr}
           \sl Diagrams contributing to $f_A(x_0)$ and $f_P(x_0)$ at
           one loop order of perturbation theory. The dotted lines
           denote the links between $x_0=0$ and $x_0=a$.}
\end{figure}

At 1--loop order, $f_A$ and $f_P$ now get several contributions:
\begin{enumerate}
\item
contributions from the first and second order terms of the link variables
$U(z-a\hat{0},0)$ and $U(y-a\hat{0},0)^{-1}$, resulting in
diagrams~1a, 1b and~2 of figure~\ref{fig:diagr},
\item
contributions from contractions between the first order terms of the link
variables and the first order terms of the quark propagators, resulting
in diagrams~3a, 3b, 4a and~4b of figure~\ref{fig:diagr},
\item
contributions from the second order terms of the quark propagators, resulting
in diagrams~5a, 5b, 6a and~6b of figure~\ref{fig:diagr}, and
\item
one contribution from the contraction of the first order terms of both
quark propagators, resulting in diagram~7 of figure~\ref{fig:diagr}.
\end{enumerate}

\begin{figure}
  \noindent
  \begin{center}
  \begin{minipage}[b]{.3\linewidth}
     \centering\epsfig{figure=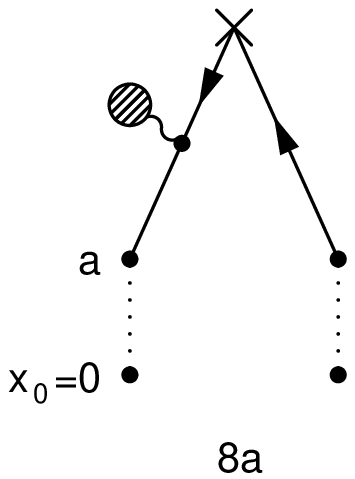,width=.8\linewidth}
  \end{minipage}
  \begin{minipage}[b]{.3\linewidth}
     \centering\epsfig{figure=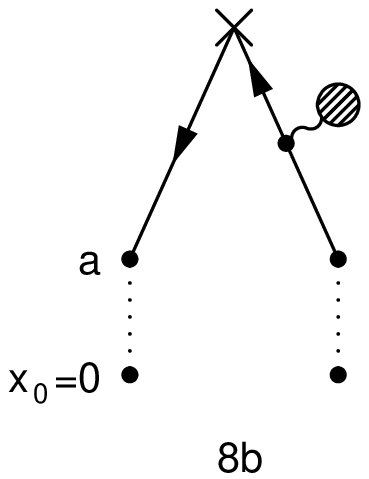,width=.8\linewidth}
  \end{minipage}\\
  \begin{minipage}[b]{.3\linewidth}
     \centering\epsfig{figure=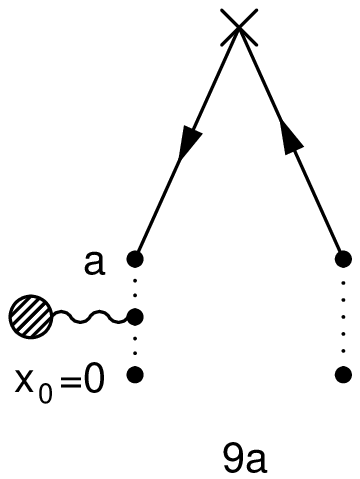,width=.9\linewidth}
  \end{minipage}\hspace{4mm}
  \begin{minipage}[b]{.3\linewidth}
     \centering\epsfig{figure=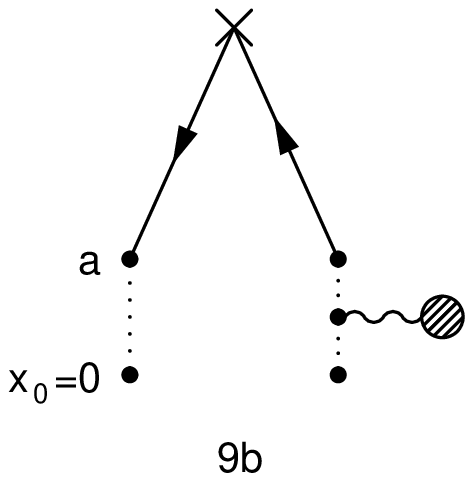,width=.9\linewidth}
  \end{minipage}\\
  \vspace{4mm}
  \begin{minipage}[b]{.6\linewidth}
     \centering\epsfig{figure=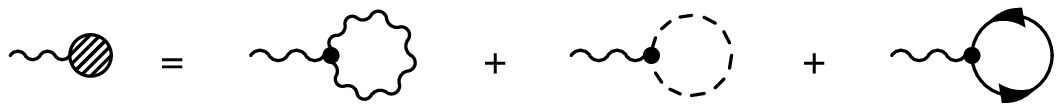,width=\linewidth}
  \end{minipage}
  \end{center}
  \caption{\label{fig:tad} 
           \sl Tadpole diagrams contributing to $f_A(x_0)$ and $f_P(x_0)$
           at one loop order of perturbation theory with non--vanishing
           background field. The dashed line represents the ghost propagator.}
\end{figure}

As already mentioned in section~\ref{sec:latticetheory}, we need a non-vanishing
background field in order to define the running coupling. Due to the presence
of this field, we obtain four more contributions. 
Taking the average $\langle\ldots\rangle_G$ leads not only to the diagrams in 
figure~\ref{fig:diagr}, but also to contractions of the first order term of the
link variables and quark propagators with the first order term of the total
action, including the gauge fixing and ghost terms needed 
for perturbation theory (see~\cite{heplat9207009,heplat9309005}). With zero
background field, this first order term vanishes, but here
it has to be taken into account. These
contributions result in the diagrams of figure~\ref{fig:tad}.
Note that only the diagrams containing closed fermion loops depend on the
number of flavours $\Nf$. So the only $\Nf$ dependent contributions to
$f_A$ and $f_P$ at 1--loop order come from diagrams~8a and~8b, whereas diagrams~9a
and~9b are of opposite sign and thus cancel in the sum. Thus $m_c^{(1)}$ becomes
$\Nf$ dependent,
\beqn
m_c^{(1)} = m_{c0}^{(1)} + m_{c1}^{(1)}\Nf.
\eeqn

In contrast to the case of vanishing background field, the propagators are not
known analytically, so they have to be computed numerically. Here, we have used
the method described in~\cite{narayanan:1995}. Due to these 
numerical computations,
computer time is not negligible. For example, on a 200 MHz Pentium PC, 
the computation of
$f_A$ and $f_P$ at $L/a=$ 16
at 1--loop level took us about 16 hours of CPU time. As one has to sum over
three momentum components and two vertex times, the time needed scales
asymptotically with $(L/a)^5$.

The 1--loop correlation functions $f_A^{(1)}$ and $f_P^{(1)}$ are
found by summing all the diagrams. We are now able to compute $m_c^{(1)}$
and thus get the deviation $\delta(u,a/L)$ up to 2--loop order. The 
results are shown in tables~\ref{tab:pertresults} and~\ref{tab:pertmass}. 
As stated before, the 2--loop coefficients
$\delta_{2j}$ are found to be small.

\section{Monte Carlo simulations with Bermions}

\subsection{The bermion model}

In the following, we will especially be interested in the theory in
which the number of flavours has been continued to
$\Nf=-2$~\cite{Anthony:1982fe,heplat9507020,heplat9907007}. This has
to be done after the integration over the fermion fields has been
performed. At $\Nf=-2$ the partition function
\beqn
  Z = \int D[U] e^{-S_g} \det(D^\dagger D)^{\Nf/2}
\eeqn
can be written as
\beqn
  Z = \int D[U] D[\phi^+] D[\phi] e^{-S_g-S_b}
\eeqn
with a local bosonic action 
\beqn
  S_b[U,\phi] = a^4\sum_x \vert(D\phi)(x)\vert^2.
\eeqn
Note that, in the actual numerical simulation, we have used the hopping
parameter representation $M$ of the Dirac operator, which is related to
$D$ by
\beqn
  M = 2\kappa(D+m_0),\qquad \kappa = (8+2am_0)^{-1}.
\eeqn
In a similar way, we can express expectation values of fermionic
observables at $\Nf=-2$ after integration over the quark fields by
expectation values in the bermion theory.  As an a priori guess
consistent with 2-loop perturbation theory, we chose the improvement
coefficient $\csw$ by linearly extrapolating the non perturbative
results at $\Nf=2$ and $\Nf=0$,
\beq
\label{eq:csw}
  \csw(g_0)\bigr\vert_{\Nf=-2} = 2 \csw(g_0)\big|_{\Nf=0} 
                                 - \csw(g_0)\big|_{\Nf=2}.
\eeq
This choice guarantees that the observables are smooth functions of
the bare coupling and an extrapolation to the continuum limit is
feasible. Furthermore, we have also computed the value for $\csw$ for
the most critical parameters used in this work along the lines
of~\cite{heplat9803017} and found good agreement with~(\ref{eq:csw}),
see the next subsection.  For the other improvement coefficients, we
take the perturbative results with their explicit respective $\Nf$
dependence.

In the bermion model, the occurrence of Dirac operator zero
modes is dynamically enhanced. Thus, in situations in which zero modes
(or exceptional configurations) are to be expected, such as large
physical volumes or large values for the bare coupling, these 
may render the functional integral ill-defined or at least hamper its
Monte Carlo evaluation.
However, we will study this theory
not too far from the perturbative regime so that these problems do not
occur. Already in the quenched approximation, exceptional
configurations occur and invalidate measured fermionic correlation functions. 
To reach larger couplings, one could also here consider a twisted mass
term~\cite{Frezzotti:1999vv,Frezzotti:2001nk}, which we shall however not pursue
in this paper.

\subsection{The size of $\csw$}
\label{sec:check}

In perturbation theory, $\csw$ is linear in $\Nf$ up to two loops,
which motivated us to first do a linear extrapolation of existing
non-perturbative data. To corroborate this further, we have computed
$\csw$ also non-perturbatively at $\Nf=-2$ for the bare coupling $\beta=8.99$,
which was used in our simulations, compare
table~\ref{tab:res_improved_L}. The computation was done along the
lines of ref~\cite{heplat9803017}, to which we refer for 
unexplained notation and
an explanation of the method.

We have computed the current mass $aM$ and the lattice
artefact $a\Delta M$ for various values of $\kappa$ at three trial
values of $\csw$.
As seen before in~\cite{heplat9609035} and~\cite{heplat9709022}, 
we note that $\Delta M$ depends only weakly on the mass $M$. 
Therefore, we are satisfied with a current mass $M$
that roughly vanishes, thereby introducing a negligible error. 
Our results for the lattice artefact are summarized in
table~\ref{tab:csw2}.  

\begin{table}[htbp]
  \begin{center}
    \begin{tabular}{lrr}
      \hline\hline
      $\csw$ & \multicolumn{1}{l}{$aM$} & \multicolumn{1}{l}{$a\Delta M$} \\
      \hline
      1.171815 &  0.0095(1) &  0.00218(19) \\
      1.271815 & -0.0003(2) &  0.00066(24) \\
      1.371815 &   0.0006(2) & -0.00127(21) \\
      \hline\hline
    \end{tabular}
    \caption{\sl $a\Delta M$ at three trial values of $\csw$ at $\beta=8.99$.}
    \label{tab:csw2}
  \end{center}
\end{table}

A linear interpolation of these three points to the improvement
condition $a\Delta M = 0.000277$ yields $\csw=1.285(7)$, which is to be
compared with the value $\csw=1.271815$ used in the simulation. 
The effect of this difference in $\csw$ on the step scaling function
can be estimated at 1-loop order of perturbation theory. For all
lattice sizes, the change in $\Sigma(u,a/L)$ is smaller than $4\times
10^{-4}$, which is negligible compared to the statistical errors,
compare table~\ref{tab:res_improved_2L}.

\subsection{Simulation algorithm}

Unimproved Wilson bermions can be simulated with a hybrid
overrelaxation algorithm, in which the gauge fields are generated by a
combination of heatbath and overrelaxation steps and the bermion
fields are generated by overrelaxation
steps only~\cite{heplat9907007}. Since the improved bosonic action depends 
quadratically on an individual gauge link $U(x,\mu)$ through the clover term,
we found it practically
impossible to generalize finite step size algorithms to
simulations of improved bermions. 
A local (link by link) hybrid Monte Carlo algorithm would be feasible and experience
from pure gauge theory shows that it is worthwhile to consider it~\cite{Gehrmann:1999wr}.
However, for the same reason as above, a part of the force would have to be
recomputed at each step on the trajectory. Thus we expect that such an
algorithm would be very expensive. Also a global hybrid Monte Carlo
algorithm would be expensive, of course. 

Therefore, we decided to perform global heatbath steps for the bosonic
field and local overrelaxation steps with respect to the unimproved
action for the gauge fields. The clover term is then taken into
account in an acceptance step in which the local action difference
with respect to the full improved action is used. The acceptance rate
in this step turns out to be large enough to pursue this algorithm.
As the overrelaxation step can be set up symmetrically, the combined
update fulfils detailed balance. Together with the heatbath step for
the bosonic field, our algorithm also satisfies ergodicity.

In principle, boson fields $\phi'$ with the
correct distribution can be generated by drawing a random field $\eta$
from a Gaussian distribution and then applying $\phi'=M^{-1}\eta$.
However, this procedure is expensive because it requires to run a
solver with full accuracy for each update. A method published
in~\cite{condmat9811025}, which is based on an approximate inversion
followed by an additional acceptance step, allows to reduce the cost
of this step.

For the update of the gauge fields, we perform a sweep over the
lattice and update the links sequentially. By an overrelaxation step with
respect to the unimproved action we propose a new configuration $U'$
which differs from the old configuration only for the link variable
$U(x,\mu) \rightarrow U'(x,\mu)$. Thus, for the acceptance step, only
the part of the action that depends on this link variable is needed.
For the gauge part of the action, the difference \beq S_g[U'] - S_g[U]
= - \frac \beta 3 \Re\Tr \Big( \big(U'(x,\mu)-U(x,\mu)\big)
S^\dagger(x,\mu) \Big) \eeq can easily be obtained. Here,
$S^\dagger(x,\mu)$ is the sum of the staples at $(x,\mu)$ and
$\beta=6/g_0^2$.

The bermion contribution can also be computed as a local difference.
At the beginning of a sweep through the lattice, the Dirac operator is
applied on the whole lattice, and the auxiliary field
$\psi=M[U]\phi$ is stored.  In each local step, $\psi'=M[U']\phi$ is computed
from $\psi$ by modifying it at those lattice sites that depend on the
link variable $U(x,\mu)$ either through the hopping or through the
clover term. In order to simplify our notation, we split $M = M_1 +
M_2$ into a term $M_1$ which is diagonal in coordinate space and a
term $M_2$ which contains nearest-neighbour contributions.
Then we need to compute
\beqn
  \Delta^{(1)}_{x\mu} \phi(z) &=& M_1[U']\phi(z) - M_1[U]\phi(z), \nonumber\\
  \Delta^{(2)}_{x\mu} \phi(z) &=& M_2[U']\phi(z) - M_2[U]\phi(z).
\eeqn
Two lattice sites are affected by the hopping term,
\beqn
  \Delta^{(2)}_{x\mu}\phi(x) &=& 
  - \kappa \lambda_\mu\ (1-\gamma_\mu)\
    \big(U'(x,\mu)-U(x,\mu)\big)\ \phi(x+a\muhat), \nonumber\\
  \Delta^{(2)}_{x\mu}\phi(x+a\muhat) &=&
  - \kappa \lambda_\mu^*\ (1+\gamma_\mu)\
    \big(U'^\dagger(x,\mu)-U^\dagger(x,\mu)\big)\ \phi(x).
\eeqn
Here $\lambda_0=1, \lambda_k=e^{i\theta}$.
Fourteen lattice sites are affected by the clover term, namely $x$,
$x+\muhat$ and for all directions $\nu\neq\mu\ $ $x\pm\nuhat$ and
$x+\muhat\pm\nuhat$. At $x$ we get for example
\beqn
  \Delta^{(1)}_{x\mu}\phi(x) &=& \frac i 8 \kappa \csw
  \big(U'(x,\mu)-U(x,\mu)\big)\times \nonumber\\
&&\times \sum_{\nu\neq\mu}
    \sigma_{\mu\nu}  
    U(x+a\muhat,\nu) U^\dagger(x+a\nuhat,\mu) U^\dagger(x,\nu),
\eeqn
and similar terms at the other points. Since the update is local, 
one has to be careful in parallelizing the algorithm. In a simulation
of a lattice on our SIMD machine with only two lattice points per
node in any direction, neighbouring nodes would modify the $\psi$ field
at a given point simultaneously through the clover term. To avoid this 
conflict, the local lattice size per node in each direction has to be larger
or equal to three.

\subsection{Performance}

As a measure for the efficiency of our algorithm, we use the machine
dependent quantity $M_{\rm cost}$ focusing on the Schr\"odinger
functional coupling $\bar g^2$. It is defined as 
\beqn
  M_{\rm cost} &=& \mbox{(update time in seconds on machine M)}\nonumber\\
 && \times (\mbox{error of $1/\bar g^2$})^2 \times (4a/T) \, (4a/L)^3.
\eeqn
In our case, $M_{\rm cost}$ refers to the CPU time spent on an 8-node
machine with APE100 architecture. This performance measure allows us
to compare for example with performance data obtained in~\cite{Frezzotti:2000rp}
for full QCD. 

A further important indicator of the efficiency of the algorithm
explained above is the acceptance rate of the clover term in
the gauge field update. This acceptance rate turns out to depend only
weakly on the parameters in the range of couplings considered here. At
$\gbar^2=0.9793$, it is about $76\%$, while at $\gbar^2=1.5145$, the acceptance is roughly
$70\%$. 

The cost of our simulations can in principle be optimized by tuning
the precision of the solver in the boson field update. We could
however only obtain a total advantage compared to a full precision
solver of roughly $10\%$ on the small lattices, with a rather flat
minimum. This can mainly be attributed to the fact that the time of an
update step is dominated by the gauge field update. Since the optimal
size of the residue has to be scaled down when increasing the lattice
size~\cite{condmat9811025}, this advantage gets even smaller on larger
lattices. Hence, we expect for our application that tuning the
precision parameter on large lattices is more expensive than running
with an ad hoc guess.

\begin{figure}[htbp]
  \begin{center}
    \vspace{-1.3cm}
    \epsfig{file=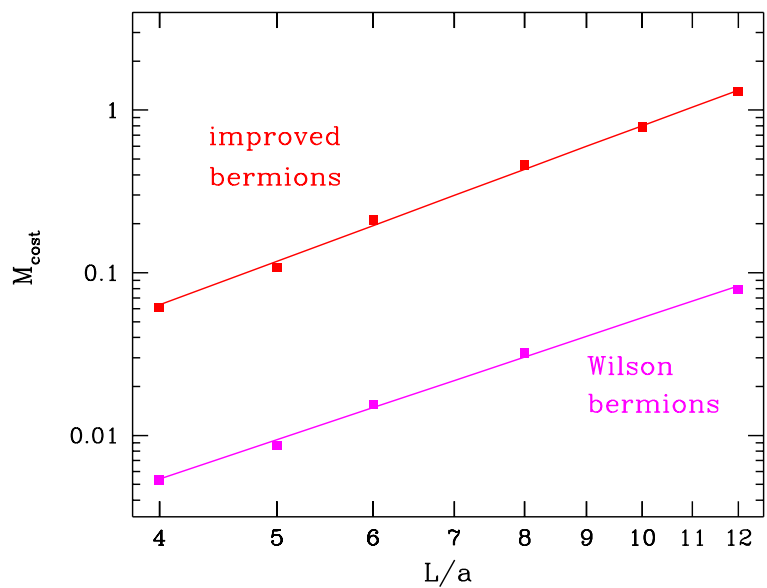,width=11cm}
    \vspace{-0.3cm}
    \caption{\label{fig:perf}
      \sl Costs for the measurement of the coupling $u=0.9793$ for
      Wilson bermions and improved bermions.}
  \end{center}
\end{figure}

The cost at $\bar g^2=0.9793$ for various lattice sizes for the
improved and the unimproved bermion theory is shown
in table~\ref{tab:perf}. 
\begin{table}[htbp]
  \begin{center}
    \begin{tabular}{rll}
      \hline\hline
      $L/a$ & $M_{\rm cost}$ & $M_{\rm cost}$ \\
            & improved &  unimproved \\
      \hline
      4  & 0.061(2)  & 0.00535(7)              \\
      5  & 0.107(3)  & 0.00866(13)              \\
      6  & 0.212(6)  & 0.0155(2)                \\
      8  & 0.457(11) & 0.0319(4)                \\
      10 & 0.790(17) &                           \\
      12 & 1.30(3)   & 0.0788(12)                \\
      \hline\hline
    \end{tabular}
    \caption{\sl Costs for improved bermions in comparison with
      Wilson bermions at $u=0.9793$. Note that the last two entries
      for improved bermions are at $u\approx 1.11$.}
    \label{tab:perf}
  \end{center}
\end{table}
Obviously, improvement of bermions (with the algorithms explained
above) leads to a substantial increase in
computer time, which we estimate to be a factor 12. In
section~\ref{sec:res}, it will become clear however, that it is still
profitable. The data of table~\ref{tab:perf} are also shown in
figure~\ref{fig:perf}.
A linear fit in this plot (that excludes the smallest lattices of the
improved theory) shows that in the improved as well
as in the unimproved theory, $M_{\rm cost}$ scales as $a^{-2.5}$.

Comparing with data from~\cite{Frezzotti:2000rp}, it turns out that
improved bermions in our implementation
are roughly a factor 10 cheaper than simulations in full QCD
with two flavours. Furthermore, the scaling with $a$ seems to be
slightly better for the bermions. On the other hand, we have estimated
the additional cost of unimproved bermions in comparison to pure gauge
theory to be only a factor 3.

\section{Technical details and results}
\label{sec:res}

\subsection{Parameters}

We have computed the step scaling function $\Sigma(u,\frac a L)$ for
the two values of the coupling $u=0.9793$ and $u=1.5145$ at lattice
sizes $L/a=4,5,6,8$. To this end, the bare parameters $\beta$ and
$\kappa$ have to be tuned such that these couplings are reached while
the current masses $m_1(L/a)$ vanish.  The precision required in the tuning
of $\kappa$ can be estimated in perturbation
theory~\cite{sommerunpublished}. In order to estimate the effect of a
slight mismatch in the tuning of $m_1$, one defines the derivative of
$\Sigma$ with respect to $z=L\ m_1$,
\beqn
  \frac \partial {\partial z} \left.\left( 
  \bar g^2(2L)\big\vert_{\bar g^2(L) = u,\ m_1(L) = z/L} 
  \right) \right|_{z=0} 
  = \Sigma'_1(a/L)\ u^2 + \ldots.
\eeqn
Under the assumption that it suffices to approximate $\Sigma'_1$ by its
universal part (valid for $L/a\rightarrow\infty$), we obtain
\beqn
  \Sigma'_1(0) = \left. 
  -\frac{\Nf}{4\pi} \frac{\partial}{\partial z} c_{1,1}(z)
  \right|_{z=0} = 0.00957\ \Nf.
\eeqn
Here, $c_{1,1}(z)$ as defined and computed in~\cite{Sint:1996ch} has
been used. This means for example that a tuning of the current mass to
zero up to $0.001$ on an $L/a=8$ lattice leads to an error in the
step scaling function smaller than $0.0002\ u^2$ and even less on the 
smaller lattices. Table~\ref{tab:res_improved_L} summarizes the
results of the tuning procedure.
These results allow to neglect the tuning error for the current mass
while the error of $\bar g^2(L)$ is propagated into the step scaling
function by perturbation theory. In those cases where $m_1=0$ is 
displayed in the table, this is a result of an interpolation in the
best tuning runs, which is justified by very small values of
$\chi^2$ obtained in the interpolation.
\begin{table}[htbp]
  \begin{center}
    \begin{tabular}{llrll}
      \hline\hline
      $\beta$ & $\kappa$ & $L/a$ & $\bar g^2(L)$ & $m_1(L/a)$ \\
      \hline
      10.3488 & 0.131024 &  4 &  0.9793(19) &  0.00000(31) \\
      10.5617 & 0.130797 &  5 &  0.9795(21) &  0.00055(13) \\
      10.7302 & 0.130686 &  6 &  0.9793(11) &  0.00000(5)  \\
      11.0026 & 0.130489 &  8 &  0.9793(14) &  0.00000(6)  \\
      \\                                       
      8.3378  & 0.132959 &  4 &  1.5145(23) &  0.00000(28) \\
      8.5453  & 0.132637 &  5 &  1.5145(17) &  0.00000(7)  \\
      8.70830 & 0.132433 &  6 &  1.5145(33) &  0.00000(4)  \\
      8.99    & 0.13209  &  8 &  1.5145(33) &  0.00066(8)  \\
      \hline\hline
    \end{tabular}
    \caption{Parameters and results for the coupling and the mass at $L$.}
    \label{tab:res_improved_L}
  \end{center}
\end{table}

\subsection{The numerical simulation}

Most of our numerical simulations were performed on APE100/Quadrics parallel
computers with SIMD architecture and single precision arithmetic. We
have used machines with up to 256 nodes with an approximate peak
performance of 50 MFlops per node. Roughly half of the statistics for
the simulation at $L/a=16$ and $u=1.5145$ has been accumulated on one crate
(128 nodes) of an APEmille installation in Zeuthen. Since our program
was not yet really optimized for APEmille, the advantage is only a factor~3.
In our simulations, we have made much use of trivial (replica)
parallelization. 

The coupling and other inexpensive observables have been measured
after each update, which corresponds to a bosonic heatbath step
followed by an overrelaxation step for the gauge fields.
The fermionic correlation functions to obtain the current mass $m_1$
have been measured only rarely, e.g.~every 100th update sweep, because
the mass does not fluctuate much. We have done up to $16\times 31500$
full updates and measurements of the coupling.
The statistical errors of the observables have been determined by a
direct computation of the autocorrelation matrix
along the lines of appendix~A of~\cite{Frezzotti:2000rp}. Typical
autocorrelation times for the coupling range from 3 to 10 (in units of
updates). 

\subsection{Discussion of results}

\begin{figure}[htbp]
  \begin{center}
    \epsfig{file=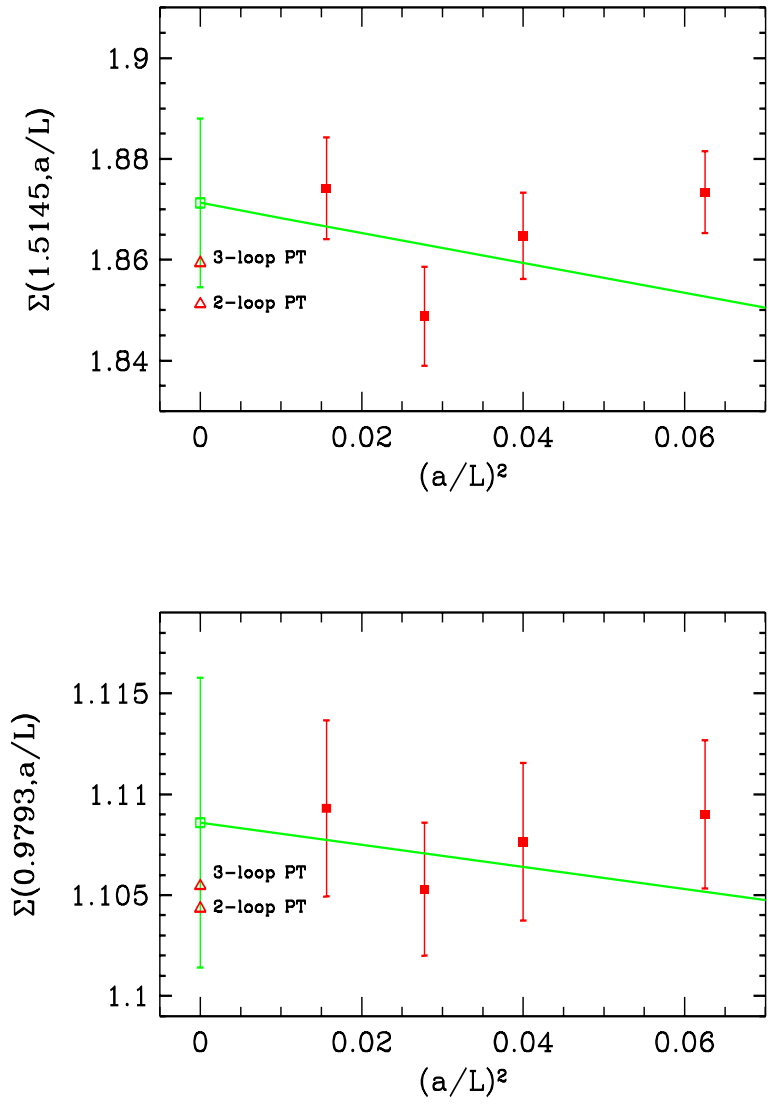,height=17.6cm,width=11cm}
    \vspace{-0.3cm}
    \caption{\label{fig:bermimp}
      \sl Step scaling function for improved bermions for the couplings 
      $u=0.9793$ and $u=1.5145$ with fits linear in $(a/L)^2$. Shown
      is also the extrapolated continuum value and the 2- and 3-loop
      results.}
  \end{center}
\end{figure}

For the non-perturbative computation of the step scaling function, we 
have simulated pairs of lattices with size $L$ and $2L$ at the same bare
parameters in the bermion theory. The results of these computations are
listed in table~\ref{tab:res_improved_2L}. 
\begin{table}[htbp]
  \begin{center}
    \begin{tabular}{lllll}
      \hline\hline
      $L/a$ & $\bar g^2(L)$ & $\bar g^2(2L)$ & $m_1(2L/a)$ \\
      \hline
      4 &  0.9793(19) &  1.1090(28) &  -0.00300(10) \\
      5 &  0.9795(21) &  1.1079(29) &  -0.00086(5)  \\
      6 &  0.9793(11) &  1.1053(30) &  -0.00094(4)  \\
      8 &  0.9793(14) &  1.1093(40) &  -0.00025(3)  \\
      \hline
      4 &  1.5145(23) &  1.8734(74) &  -0.00266(12) \\
      5 &  1.5145(17) &  1.8648(82) &  -0.00094(7)  \\
      6 &  1.5145(33) &  1.8488(86) &  -0.00070(5)  \\
      8 &  1.5145(33) &  1.869(14)  & $\,\!$ 0.00002(5)  \\
      \hline\hline
    \end{tabular}
    \caption{\sl Results for the coupling and the mass at $2L$ at
      the parameters defined by the given value of $\bar g^2(L)$.}
    \label{tab:res_improved_2L}
  \end{center}
\end{table}
For the propagation of the statistical error and the mismatch of the
tuning results for the coupling, we use a perturbative ansatz. Then we
obtain the lattice step scaling function $\Sigma(u,a/L)$ that is
shown in figure~\ref{fig:bermimp}.
We pass to the continuum limit by an extrapolation in $a/L$ with the
ansatz
\beqn
  \Sigma(u,a/L) = \sigma(u)\ (1 + \rho(u) (a/L)^2).
\eeqn
As shown in figure~\ref{fig:bermimp}, this ansatz works
perfectly, i.e.~within the error bars no linear dependence of the step
scaling function on $a/L$ can be detected. The results for the
continuum step scaling function and the corresponding 2- and 3-loop
values are given in table~\ref{tab:r2}.
\begin{table}[htbp]
  \begin{center}
    \begin{tabular}{llll}\hline\hline
      $u$ & $\sigma(u)$ & 
        $\sigma(u)\vert_{\mbox{\footnotesize 2-loop}}$ & 
        $\sigma(u)\vert_{\mbox{\footnotesize 3-loop}}$ \\ 
     \hline
     0.9793 & 1.1063(46) & 1.10435 & 1.10691 \\ 
     1.5145 & 1.871(17)  & 1.85122 & 1.87026 \\
     \hline\hline
     \end{tabular}
     \caption{\sl Extrapolated simulation results and perturbation theory for the
       step scaling function at $\Nf=-2$.} 
    \label{tab:r2}
  \end{center}
\end{table}
The difference between 2- and 3-loop perturbation theory is thus of
the same size as the error of the extrapolated simulation
results. Within the error bars, both values of the step scaling function
$\sigma(u)$ are consistent with perturbation theory.

\begin{figure}[htbp]
  \begin{center}
    \epsfig{file=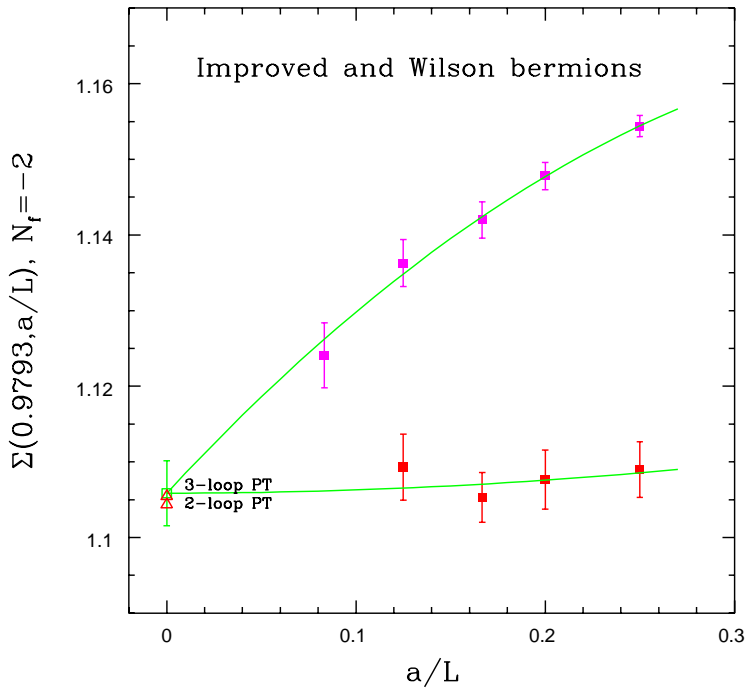,height=11cm,width=11cm}
    \vspace{-0.3cm}
    \caption{\label{fig:bermcombined}
      \sl Results for the step scaling function at $u=0.9793$, together
      with a quadratic fit under the constraint of universality.}
  \end{center}
\end{figure}

The typical size of the $\rmO(a)$ effects can be estimated by
considering the step scaling function in the unimproved bermion theory
as well. This has been done 
in~\cite{heplat9907007} for $u=0.9793$. Figure~\ref{fig:bermcombined}
shows these results together with the data after implementing improvement. 
Obviously, the
linear cutoff effects are quite seizable for this observable, of the
order of a few percent. The data are
fitted with a combined fit under the constraint that their continuum
limit agrees, that means assuming universality. This fit is linear
plus quadratic in $a/L$ for the Wilson bermion data and quadratic in
$a/L$ in the improved data. Although the additional input from the
Wilson data is included, the joint continuum limit $\sigma_{\rm
  combined}(0.9793)=1.1059(43)$ agrees almost completely with the
value in table~\ref{tab:r2}. A linear plus quadratic fit in $a/L$ of
the unimproved data alone would have given the continuum result
$\sigma_{\rm unimproved}(0.9793) = 1.103(12)$.

The uncertainties of these extrapolated values show a remarkable
success of improvement. Although the total computer time for the
improved simulations was only by a factor $1.7$ higher than for the
Wilson bermions, the error after extrapolation is by a factor $2.7$
smaller. Since a large portion of the computational cost for
$\sigma(u)$ comes from the largest lattice, this advantage can be
attributed to the lattice size needed for a reasonable extrapolation.
For Wilson bermions, this was $L/a=24$, whereas simulations are 
limited to $L/a=16$ for the improved case. Of course, our
observation is restricted to one value of the coupling; at different
values, lattice artefacts may behave differently. It should also be
noted that in the bermion case, adding the clover term leads to a
significant performance penalty for the numerical simulation. For
typical algorithms for the simulation of dynamical fermions, like
variants of the Hybrid Monte Carlo, the inclusion of the clover term
implies a much smaller overhead. Hence, the advantage of improvement
should be even bigger there.

\begin{figure}[htbp]
  \begin{center}
    \epsfig{file=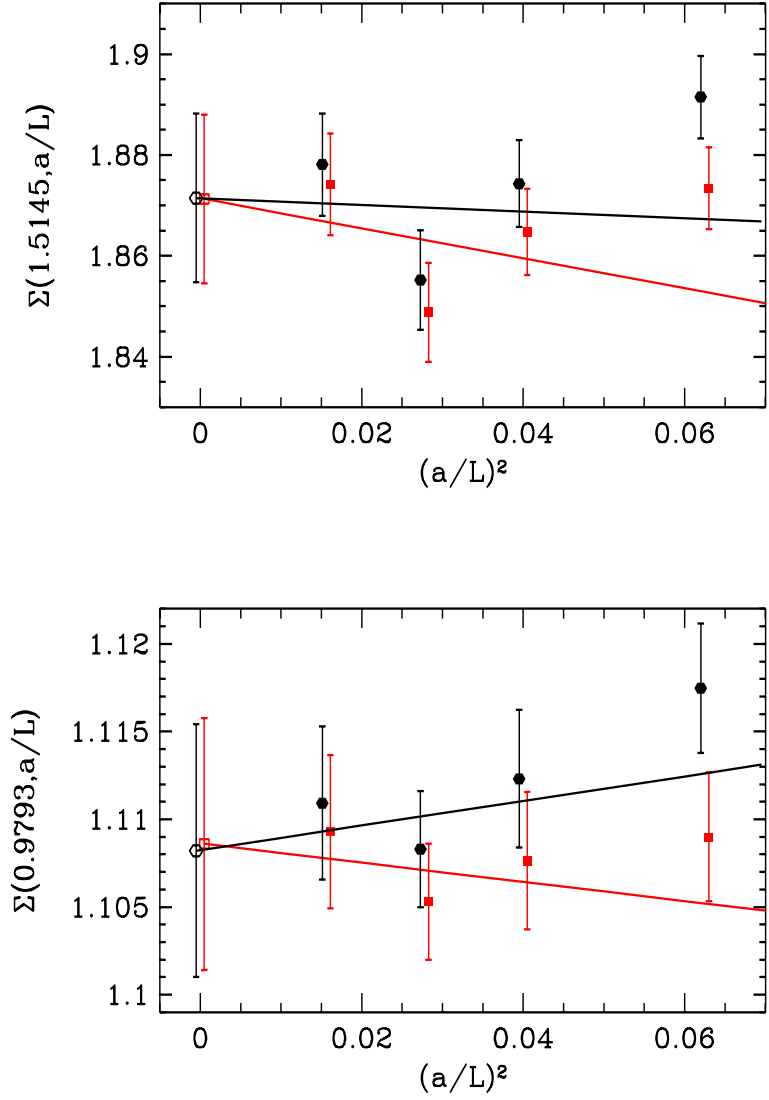,height=17.6cm,width=11cm}
    \vspace{-0.3cm}
    \caption{\label{fig:bermcorrected}
      \sl Step scaling function for improved bermions for the couplings 
      $u=0.9793$ and $u=1.5145$ with fits linear in $(a/L)^2$. The 
      rectangles represent the data points obtained from our simulations,
      whereas the circles represent the data $\Sigma^{(2)}$
      corrected by perturbation theory.}
  \end{center}
\end{figure}

In addition to our determination of the step scaling function by
extrapolating $\Sigma(u,L/a)$ from Monte Carlo simulations, we have also
analysed the approach to the continuum limit for data which have
the 2-loop perturbative lattice artefacts cancelled.
To this end, we replace the lattice step scaling used for the
fit by the corrected values
\beq
  \Sigma^{(2)}(u,a/L) 
  = \frac{\Sigma(u,a/L)}{1+\delta_1(a/L)u + \delta_2(a/L)u^2}
\eeq
with $\delta$ from ~(\ref{eq:delta}).
The results for these fits are shown in figure~\ref{fig:bermcorrected},
together with the uncorrected fits shown before in
figure~\ref{fig:bermimp}. Again, we leave out the point at $L/a=4$
for the fits. As can be seen from the plot, this procedure does
not visibly reduce remaining lattice artefacts. For the coupling
$u=1.5145$, the slope of the fitted line gets slightly smaller,
whereas for $u=0.9793$ it remains roughly equal. However, if we
also include the point at $L/a=4$, the lattice artefacts have even larger
$\rmO(a)^2$ effects than for the uncorrected data.
Nevertheless, their continuum limit agrees within the error bars
with the one obtained by the procedure used before. It is also
consistent with perturbation theory in the continuum limit.

\begin{figure}[htbp]
  \begin{center}
    \epsfig{file=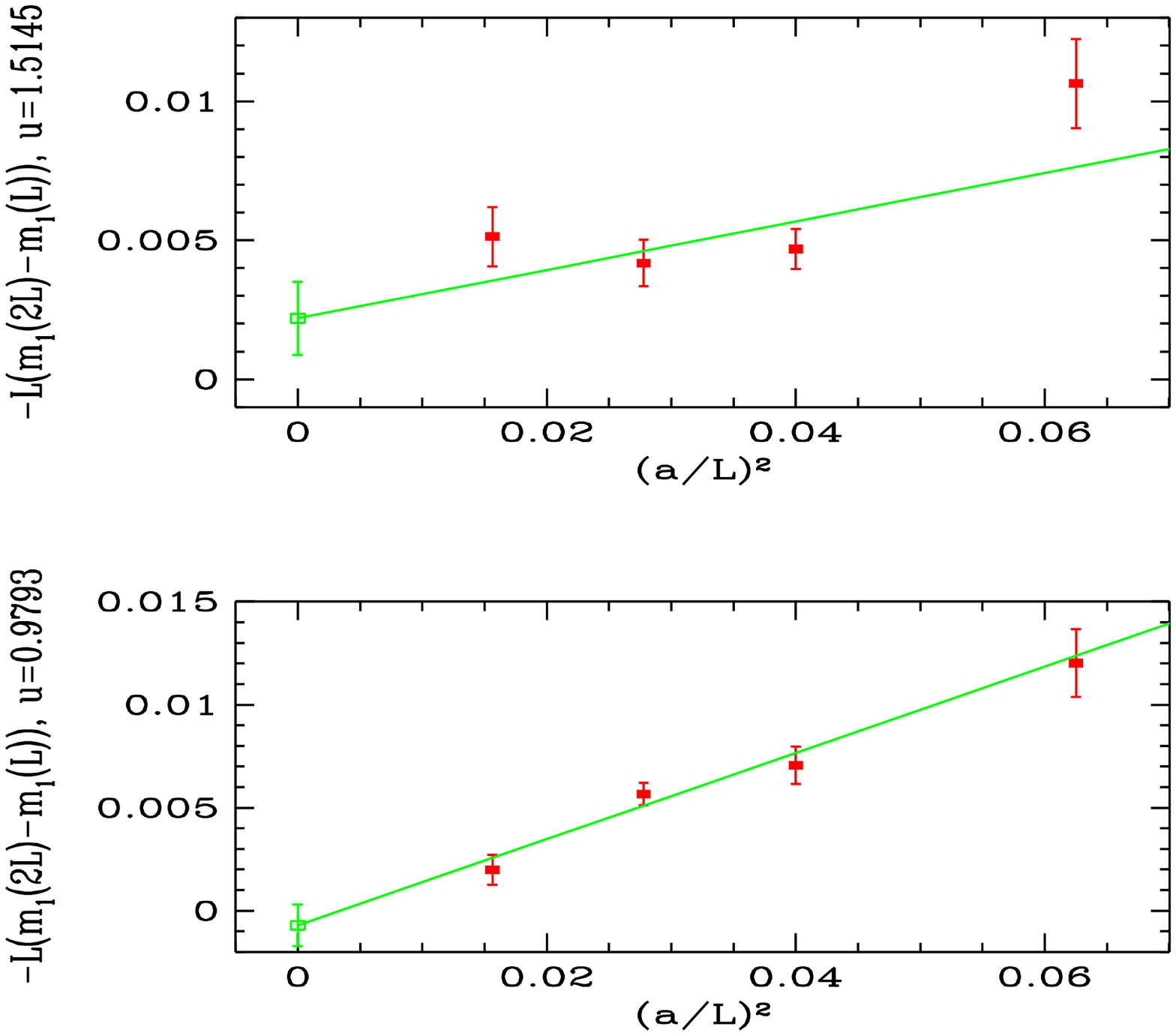,height=17.6cm,width=11cm}
    \vspace{-0.3cm}
    \caption{\label{fig:masses}
      \sl Check of lattice artefacts in the mass at $u=0.9793$ and $u=1.5145$.}
  \end{center}
\end{figure}

Since the mass $m_1$ is tuned to zero on the small lattices, we expect
that $L\big(m_1(2L)-m_1(L)\big)$ is a pure lattice artefact that vanishes in
the continuum limit with a rate proportional to $(a/L)^2$. This
expectation is confirmed by figure~\ref{fig:masses} in which 
this mass difference is shown as a function of $(a/L)^2$. While the
scaling is perfect for the smaller of the two couplings, there are
small deviations at $u=1.5145$, which however can be attributed to
statistical fluctuations.

\section{Conclusions}

In this paper, we have investigated the step scaling function in the
$\rmO(a)$ improved bermion model by means of extrapolating Monte Carlo 
data at different lattice resolutions to the continuum limit. 
The  results obtained were compared with renormalized perturbation
theory in the continuum limit and with data obtained from simulation
of unimproved bermions. They also serve as a guide in planning
analogous simulations with dynamical quarks. 

It is demonstrated that the implementation of improvement successfully
reduces lattice artefacts and allows a fit of the step scaling
function $\Sigma(u,a/L)$ linearly in $(a/L)^2$ (see figure~5).  This raises our
confidence that the same extrapolation procedure can be applied for
dynamical fermions. We note however, that to our disappointment, the
computer time required for our algorithm turned out to be only about a
factor $10$ smaller than for dynamical fermions. This means that the
lattice sizes that can be reached are not much larger than for
fermions, in contrast to the situation without improvement.

\vskip 1cm
\noindent
{\bf Acknowledgements.}
We would like to thank Peter Weisz for essential checks
on our perturbative calculations and Rainer Sommer for 
helpful discussions.
DESY provided us with the necessary computing resources and the APE 
group contributed their permanent assistance.
This work is supported
by the Deutsche Forschungsgemeinschaft under 
Graduiertenkolleg GK 271 and by the
European Community's Human Potential Programme
under contract  HPRN-CT-2000-00145.

\bibliographystyle{myunsrt}
\bibliography{lit}

\end{document}